\documentclass[12pt]{article}
\usepackage{graphicx}

\date{}
\title{\bf 
Dynamics Of Proton Spin :  
 Role Of  $qqq$  Force }
\author{
{\normalsize\bf
A.N.Mitra \thanks{Email: (1)ganmitra@nde.vsnl.net.in;
(2)anmitra@physics.du.ac.in}
}\\
\normalsize 244 Tagore Park, Delhi-110009, India}
       
\begin{document}

\maketitle

\begin{abstract}
The analytic structure of the  $qqq$ wave function, obtained  recently  in the high momentum regime of QCD,  is employed for the formulation of baryonic transition amplitudes via quark loops.   A new aspect  of this study is the role  of a  direct ( $Y$-shaped, Mercedes-Benz type) $qqq$  force  in generating  the $qqq$ wave function  The dynamics  is that of a  Salpeter-like equation  ( 3D support for the kernel)  formulated covariantly  on the light front, a la   Markov-Yukawa Transversality Principle (MYTP)  which warrants a 2-way interconnection between  the 3D and 4D Bethe-Salpeter (BSE)  forms   
for 2 as well as 3  fermion quarks. The dynamics of this 3-body force shows up through a characteristic singularity in  the hypergeometric differential equation for the 3D wave function $\phi$, corresponding to  a $negative$ eigenvalue of the spin operator  $i \sigma_1.\sigma_2\times \sigma_3$  
which is an  integral  part   of   the $qqq$ force. As a  first application of this wave function  to the problem of the proton spin anomaly, the
two-gluon contribution to the  anomaly  yields an estimate of the right sign, although somewhat smaller in magnitude.  \\
Keywords: 3bodyforce; proton-spin ; 2gluon anomaly ; fractional correction $\theta$

\end{abstract}

\section{ Introduction}

The concept  of a fundamental  3-body force (on par with a 2-body force)  is  hard to realize in physics, leaving aside certain ad hoc representations 
of higher order effects, for example  those of $\Delta, N^*$ resonances  in hadron physics. At the deeper quark-gluon level on the other hand, a  
truly 3-body $qqq$ force  shows up as a folding of a  $ggg$ vertex ( a genuine part of the gluon Lagrangian in QCD) with 3 distinct ${\bar q}gq$ 
vertices, so as to form a  $Y$-shaped diagram.  Indeed  a 3-body $qqq$ force of this type, albeit  for `scalar' gluons,  was 
first suggested  by Ernest Ma \cite{ ErMa75},  when QCD was still  in its infancy. [ A similar representation is also possible for $NNN$ interaction via 
$\rho\rho\rho$  or $\sigma\sigma\sigma$ vertices, but was never in fashion in the literature  \cite{McK94}].  We note in passing that  a 
$Y$-shaped (Mercedes-Benz type) picture \cite{Mitr83}  was once  considered in the context of a preon model for quarks and leptons. 

\setcounter{equation}{0}
\renewcommand{\theequation}{1.\arabic{equation}}

In the context of QCD  as a Yang-Mills field,  a $ggg$ vertex has a momentum representation of the form \cite{Tay78}
\begin{equation}\label{1.1}
W_{ggg} =  -i g_s f_{abc} [(k_1-k_2)_\lambda \delta_{\mu\nu} + (k_2-k_3)_\mu \delta_{\nu\lambda} + (k_3-k_1)_\nu \delta_{\lambda\mu}]
\end{equation} 
where the 4-momenta emanating from the $ggg$ vertex satisfy $k_1 + k_2 + k_3 = 0$, and $f_{abc}$ is the color factor.  When  this vertex is folded  
into 3  ${\bar q}gq$ vertices of the respective forms $ g_s {\bar u}(p_1') i\gamma_\mu \{\lambda_1^a /2 \}  u(p_1)$, and two similar terms,  the 
resultant $qqq$ interaction matrix (suppressing the Dirac spinors for the 3 quarks)  becomes \cite {Mitr07}
\begin{equation}\label{1.2} 
V_{qqq} = \frac{ g_s^4}{2^3} [ i\gamma^{(1)}. (k_2 - k_3)  \gamma^{(2)}. {\gamma^{(3)}  + \{2\} + \{3\} ]  \{\lambda_1\lambda_2\lambda_3\}  /\{ k_1^2 k_2^2 k_3^2} \} 
\end{equation} 
where $k_i = p_i -p_i' $;  $\lambda_i$   are the color  matrices  which get contracted into the 
corresponding scalar triple products  in an obvious notation. [Note that   the  flavour indices are absent  here since the quark gluon interaction is 
flavour blind].   
\par    
This interaction will  be considered in conjunction  with 3  pairs of $qq$ forces  within the framework of a Bethe-Salpeter type dynamics to be specified below.  Before proceeding  further, a possible motivation  for the use of  a direct $qqq$ force, apart from its intrinsic beauty,  comes  from  the  issue of "proton spin"   which,  after making  headlines about two decades ago, has  come to the fore once again, thanks to the progress of  experimental  techniques in polarized deep inelastic scattering off polarized protons, and their variations thereof,  which allow for an experimental  determination of  certain key  QCD  parameters by relating them  to certain  observable quantities emanating from  external probes ; ( see a recent review  \cite{Bass05} for references and other details).   On the other hand it is also of considerable theoretical interest  to determine these very quantities directly  from the $intrinsic$ premises of QCD  $provided$  one has a " good " $qqq$ wave function  to play with.  Such a plea would have sounded rather utopian in the early days of QCD when phenomenology was the order of the day. Today however many aspects of QCD are understood well enough to make such studies worthwhile by hindsight, with possible ramifications beyond their  educational value. For simplicity we work in the experimentally accessible  regime of valence quarks. and make free use of the results of a recent paper \cite {Mitr07}  for many details on  the specific  effect of the 3-body force (1.2) on the analytical structure of the $qqq$ wave function, while giving more emphasis on the formalism relating to the  loop diagrams  towards the determination of  proton spin with appropriate 4D BS normalization.     

\subsection{Theoretical Ingredients}

In the valence quark regime, we need to consider a $qqq$ system governed by pairwise $qq$ forces  as well as a direct 3-quark force of the type (1.2).  
A further simplification occurs in  the high momentum regime where the effect of confining forces may be neglected, so that only coulombic forces are 
relevant. As explained in \cite{Mitr07}, we shall take  the dynamics of a $qqq$ hadron  in the $high- momentum$  regime to be  governed by   the Salpeter  equation \cite{Salp52} formulated in a covariant manner,  which  has the remarkable property of  3D-4D interlinkage ( see \cite{Mitr07} for a detailed picture).      
 A covariant formulation of the Salpeter Equation in turn,  is centered around the hadron  4-momentum $P_\mu$  in accordance with  the Markov-Yukawa Transversality Principle (MYTP) \cite{Mark40, Yuka50}, which is  a `gauge principle'  in disguise \cite{LuOz77}, and  ensures that   the interactions among the constituents  be  $transverse$ to the direction of     
$P_\mu$.   In the high momentum regime to be considered here,  the confining interaction has been ignored for simplicity \cite{Mitr07},   which leaves the     
3D form of the BS dynamics  inadequate for   mass spectral determination, yet  its dynamical effect  on the spin-structure of the wave function   should  
be realistic enough for dealing  with  the  hadron spin  in the high momentum limit.    
\par
A further ingredient concerns the use of Dirac's light-front form of dynamics \cite{Dira49} which has a bigger \{7\} stability group than  the more conventional instant form whose stability group is only \{6\}. All this can be covariantly  formulated; see \cite{Mitr07} and \cite{Mitr99} for details.

\subsection{ Plan of the Paper}  

The plan of the paper, which  is  based on  an interlinked  3D-4D BSE  formalism characterized by  a Lorentz-covariant  3D support for its kernel a la MYTP  \cite{Mark40, Yuka50}, adapted to  the light front (LF) \cite{Mitr99} is as follows.  In Section 2 we summarise the principal results of ref \cite{Mitr07} on both the 3D ($\phi$) and the full-fledged 4D($\Psi$) forms of the $qqq$ wave function, so as to give a basically self-contained picture omitting the non essential details from \cite{Mitr07}.    Section 3 outlines the construction of the normalized 4D wave function after assessing the possible options on BS normalization for the same. (As a check,  some of the conventional results are reproduced).  Section 4  is devoted to the principal result of this investigation, viz., the construction of the two-gluon coupling to the axial operator $i\gamma_\mu \gamma_5$ and its insertion into the quark lines involved in the two types ( self-energy and exchange) of possible  baryonic transition diagrams for such coupling.  Section 5 concludes with a short discussion of the results obtained vis-a-vis experiment.  

\section{ Structure of the Full BS Wave Function $\Psi$}

In this Section we collect the principal results of  ref \cite{Mitr07} on the  full structure of  the BS wave function in both the 
3D ($\phi$) and 4D ($\Psi$) forms. 
\setcounter{equation}{0}
\renewcommand{\theequation}{2.\arabic{equation}}

\subsection{Instant vs LF Representations of Momenta}

We first record the correspondence between the instant and LF forms of the dynamics, starting with   some definitions \cite{Mitr99} for the LF quantities  $p_{\pm} = p_0 \pm p_3$ defined covariantly as  $p_+ = n.p \sqrt{2}$ and $p_- = -{\tilde n}.p\sqrt{2}$. while the perpendicular components   continue to be denoted by $p_\perp$ in both notations.  For  a typical  internal momentum $q_\mu$,  the  parallel component  $P.q P_\mu /P^2$  of the instant form  translates in the LF form  as  $q_{3\mu} = z P_n n_\mu $, where $P_n = P.{\tilde n} $, and $z = n.q / n.P$. As a check,      
${\hat q}^2 = q_\perp^2 + z^2 M^2$ which shows that $zM$ plays the role of the third component of ${\hat q}$ on LF. Next,  we collect   some of the    
more important   definitions / results of the LF formalism  \cite{Mitr99} 
\begin{eqnarray}
q_\perp & = & q - q_n n ;   {\hat q} = q_\perp + z P_n n;  z  =  q.n / P.n ;   q_n  =  q.{\tilde n} ;    \\  \nonumber    
 P_n  &=  &P. {\tilde n};    P.q  =  P_n q.n + P.n q_n ;  {\hat q}.{\tilde n} = P_\perp. q_\perp = 0 ;  \\  \nonumber 
P.{\hat q} & = & P_n q.n ;  {\hat q}^2 = q_\perp ^2 + M^2 z^2 ; P^2 = - M^2 
\end{eqnarray} 
For a $qqq$ baryon, there are two internal momenta, each separately  satisfying the relations (2.1). Note that for any 4-vector $A$, $A.n$ and $- A_n$ 
correspond to   $1 / \sqrt{2}$ times the usual light front quantities $A_\pm = A_0 \pm A_z$  respectively. But since a $physical$ amplitude must not depend on the orientation $n$,  a simple device termed $Lorentz-completion$ via  the $collinear$ trick \cite{Mitr99} yields a Lorentz-invariant amplitude      
for a transition process  with  $three$ external lines $P, P', P'' (= P + P') $ as  explained in \cite{Mitr99, Mitr07}. And for ready reference,  the precise correspondence between the  instant and LF  definitions of the  `parallel (z)'  and `time-like (0) ' components of the     
various 4- momenta for a $qqq$ baryon ( i = 1,2,3) \cite{ MitS01} : 
\begin{equation}\label{2.2}
p_{iz}; p_{i0} = \frac{M p_{i+}}{P_+}; \frac{M p_{i-}}{2 P_-} ; \quad {\hat  p}_i  \equiv  \{ p_{i\perp} , p_{iz}\}
\end{equation}
The last part of Eq.(2.2)  defines a covariant  3-vector on the LF that will frequently  apper as arguments of  3D wave function $\phi$ for the $qqq$ proton.    

\subsection{ From  $\Psi$ to $\Phi$ via Gordon Reduction }

The full wave function  for  three fermion quarks complete with all internal d.o.f.'s,  satisfies the following Master equation  whose kernel includes 
both $qq$ and  direct $qqq$ forces   \cite{CGSM89} : 
\begin{eqnarray}\label{2.3}
\Psi (p_1p_2p_3) &=&  \sum_1^3 S_F(p_1) S_F(p_2) g_s^2 \int \frac{d^4 q_{12}'}{(2 \pi)^4} \gamma_\mu^{(1)} \gamma_\nu^{(2)}
  D_{\mu\nu} (k_{12}) \Psi(p_1', p_2', p_3)  \nonumber  \\ 
                            &  &  +  S_F(p_1) S_F(p_2) S_F(p_3) \int \frac{d^4 q_{12}' d^4 p_3' }{(2 \pi)^8 }  V_{qqq} \Psi(p_1' p_2' p_3')
\end{eqnarray} 
where  the definitions  for the various momenta,  and the phase conventions for the quark propagators  are those of  \cite{CGSM89}, while the   
direct 3-quark interaction $V_{qqq}$ in the last term is   given by (1.2). Here the internal variables must be defined in a pre-assigned basis, say   
 indexed by  $\#3$  as \cite{MitS01} 
\begin{equation}\label{2.4}
\sqrt{2} \xi_3 = p_1-p_2 ; \quad \sqrt{6} \eta_3 = -2p_3+p_1+p_2; \quad
P=p_1+p_2+p_3 
\end{equation}
where the  time-like and space- like parts of each are given by (2.2), and the corresponding 3-vector defined as  ${\hat p}_i \equiv \{p_{i\perp}, p_{iz}\}$.  
(Two identical sets of momentum pairs $\xi_1, \eta_1$ and $\xi_2, \eta_2$ are similarly defined, but can be expressed in terms of the set (2.4) 
via permutation symmetry).  The solution of this Master equation (2.3) was then achieved in  three steps (A, B, C). 
Step A consists in  defining  an auxiliary scalar  function $\Phi$ related to the actual BS wave function $\Psi$ by \cite{CGSM89} 
\begin{equation}\label{2.5}
\Psi = \Pi_{123} S_{Fi}^{-1}(-p_i) \Phi (p_i p_2 p_3) W(P)  
\end{equation}
where the quantity $W(P)$ is independent of the internal momenta but includes the  spin-cum-flavour wave functions $\chi, \phi$ of the 3 quarks 
involved  (see  \cite{MiMi84}  for notation and other details) :      
\begin{equation}\label{2.6}
W(P) = [ \chi' \phi' + \chi'' \phi'' ] / \sqrt{2}
\end{equation}
The quantities $\phi'$, $ \phi''$ are the standard  flavour functions of mixed symmetry \cite{Feyn71} [not to be confused with the 3D wave function $\phi$ !], 
and $\chi'$ ,$ \chi''$ are the corresponding {\it relativistic}  spin functions. The latter may be  defined either in terms of the quark \# indices as in Eqs (1.2)  
or (2.3), or sometimes more conveniently in a common Dirac matrix space as \cite{MiMi84, Blan59} 
\begin{equation}\label{2.7}
|\chi'> ; |\chi''> = [\frac{M - i\gamma.P}{2 M}[i\gamma_5; i{\hat \gamma}_\mu / \sqrt{3}] C / \sqrt{2}]_{\beta\gamma} \otimes 
[[1; \gamma_5{\hat \gamma}_\mu] u(P)]_\alpha
\end{equation} 
where the first factor is the $\beta \gamma$-element of a 4 x 4 matrix in the joint spin space of the  quark \#s 1, 2  \cite{Blan59}, and the second factor      
the $\alpha$ element of a 4 x 1 spinor in the spin space of quark \# 3; $C$ is a charge conjugation matrix with the properties \cite{Davi65}
$$ - {\tilde \gamma}_\mu = C^{-1}\gamma_\mu C ;  {\tilde \gamma}_5 = C^{-1} \gamma_5 C ; $$
and ${\hat \gamma}_\mu$ is the component of $\gamma_\mu$ orthogonal to $P_\mu$. Finally, the representations of the flavour functions $\phi', \phi''$
satisfy the following relations in the "3" basis \cite{MiRo67}  
\begin{equation}\label{2.8}
 < \phi'' | 1 ; {\vec \tau}^{(3)} | \phi'' > =  < \phi' | 1 ; - \frac{1}{3} {\vec \tau}^{(3)}  | \phi' >  
\end{equation}
Step B now consists in recasting Eq.(2.3) in terms of the scalar quantity $\Phi$  a la  Eq.(2.5) with a simultaneous use of Gordon reduction  on the pairwise kernels $V({\hat \xi}_i{\hat \eta}_i)$ and the 3-body kernel $V_{qqq}$,  as described in \cite{Mitr07} following  the original treatment of  \cite{Mitr81}.  This 
has the effect of eliminating the Dirac matrices in favour of the  Pauli matrices $\sigma_{\mu\nu}$. [We skip these details which may be found in 
\cite{Mitr07}]. 

\subsection{ 3D-4D Interlinkage by  Green's Function  Method} 

The next step (Step C) now consists in a  reduction of the 4D BSE for the  quantity $\Phi$ defined above  to one for a 3D scalar $\phi$ by the standard method of elimination of the time-like variables, and a reconstruction of the 4D quantity $\Phi$ , thus  establishing a 3D-4D interconnection  between 
these two  wave functions. This last  is facilitated by  the Green's function approach \cite{MitS01} adapted to the LF formalism, as described in \cite{Mitr07}.   Calling the 4D Green's functions associated with 
$\Psi$ and $\Phi$ by $G_F$ and $G_S$ respectively, the  connection between them, analogously to Eq.(2.5),  may be written as 
\begin{equation}\label{2.9}
G_F (\xi\eta;\xi'\eta')=  W(P) \otimes  { \Pi_{123} S_{Fi}^{-1}(-p_i)} G_S (\xi\eta;\xi'\eta') {\Pi_{123} S_{Fi}^{-1}(-p_i')} {\bar W}(P') 
\end{equation} 
where we have indicated the 4-momentum arguments of the Green's functions involved,  in a common $S_3$ basis $(\xi, \eta)$,  and expressed the 
spin-flavour dependence of $G_F$ as a  matrix product  implied by the notation $W (P) \otimes {\bar W} (P')$.  
\par
It was shown in \cite{Mitr07} how the 3D-4D interconnection is first achieved  at the level of the `scalar' Green's functions whose 4D and 3D forms are labelled by $G_S$ and $g_s$ respectively, and thence to the corresponding wave functions $\Phi$ and $\phi$ by the method of spectral representations. 
Finally the connection to the 4D spinor wave function $\Psi$ is established via Eq. (2.5).  We skip these steps which are given in sufficient details in 
\cite{Mitr07}. The final result for $\Psi$ in terms of $\phi$  is   
\begin{equation}\label{2.10}
\Psi (\xi, \eta)  = \Pi_{123} S_F(p_i) D_{123}\sum_{123} [ \phi({\hat \xi}, {\hat \eta})              
\frac{ 1}{(2\pi i)^2}] \times W(P) 
\end{equation}  
where  the structure of  $D_{123}$  is expressed by a double integral over two time-like momenta:
\begin{equation}\label{2.11}
\frac{1}{D_{123 }} =  \int \frac { P_+^2 dq_{12-} dp_{3-}}{ 4 M^2 (2i \pi)^2 \Delta_1 \Delta_2 \Delta_3} 
\end{equation}
and the  3D wave function  $\phi$ satisfies the equation      
\begin{eqnarray}\label{2.12}
(2\pi)^3 D_{123} \phi({\hat \xi}, {\hat \eta})  &=& \sum_{123} \frac{p_{3z}}{\sqrt{2}}  \int d^3{\hat \xi}_3'' V_{qq3} \phi({\hat \xi}_3'', {\hat \eta}_3)   \nonumber \\
&+& \frac{1}{3\sqrt{3} (2\pi)^3} \int d^3 \xi'' d^3 \eta'' V_{qqq} \phi({\hat \xi}_3'', {\hat \eta}_3'')
\end{eqnarray}
The solution of this equation has been obtained in \cite{Mitr07} in coordinate space, using combinations analogous to (2.4), viz.,  
\begin{equation}\label{2.13}
\sqrt{2} s_3 = r_1 - r_2 ; \quad \sqrt{6} t_3 = -2 r_3 + r_1 + r_2 
\end{equation} 
The final result for $\phi$ in coordinate space (see \cite{Mitr07} for details) is 
\begin{equation}\label{2.14}
\phi = F ( a, b | 3 | x) ; \quad a + b = 2; \quad ab = \frac{\beta}{\sqrt{2}} ; \quad \beta \approx 0.058
\end{equation}
where $F$ is a standard hypergeometric function of its arguments, and has a particularly convenient  representation for $a +b =2$ \cite{WhWa52} 
\begin{equation}\label{2.15}
F(a, b ; 3; x) = \int_0^1 dy y^{a-1} \frac{(1-y)^a}{(1-xy)^a}; \quad a \approx  2 - \beta/ \sqrt{2}
\end{equation}
where $x = R^2 / R_0^2$, $R^2 = s^2 + t^2$, and $x=1$ corresponds to the point $R = R_0$.  This completes our summary of the full structure of the 4D $qqq$ wave function $\Psi$ in terms of the 3D quantity $\phi$  a la \cite{Mitr07}.

\section{Proton Spin Formalism } 

\setcounter{equation}{0}
\renewcommand{\theequation}{3.\arabic{equation}}

As a first application of this wave function, we shall  determine  the baryon spin, together with its corrections,  in a general enough manner involving loop diagrams.  To that end a key ingredient is the baryon normalisation within the Bethe Salpeter formalism,  for which the appropriate diagram  is Fig 1 with the spin operator $i\gamma_\mu \gamma_5$ replaced by an appropriate one signifying conservation of charge, mass or probability with corresponding operators $e(1/6 + \tau / 2) i\gamma_\mu$, $M^2$ or $ 1$ respectively.   We adopt the last one (probability) in preference to the others in view of 
its simplicity as well as  universal appeal as an $even$ operator.  

 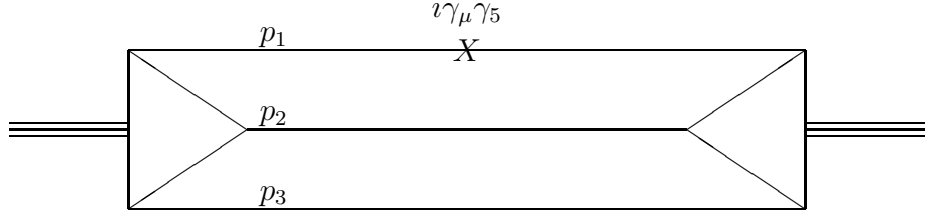
\begin{figure}[h]

\vspace{0.5in}

\begin{picture}(450,50)(-80,100)

\put (276,80){\line(0,1){60}}
\put (276,80){\line(-3,2){45}}
\put (231,110){\line(3,2){45}}

\put (20,80){\line(0,1){60}}
\put (65,110){\line(-3,2){45}}
\put (65,110){\line(-3,-2){45}}

\put (20,80){\line(1,0){256}}
\put (20,140){\line(1,0){256}}
\put (65,110){\line(1,0){166}}

\put(75,85){\makebox(0,0){$p_3$}}
\put(75,145){\makebox(0,0){$p_1$}}
\put(75,115){\makebox(0,0){$p_2$}}

\put (-25,110){\line(1,0){45}}
\put (-25,112.5){\line(1,0){45}}
\put (-25,107.5){\line(1,0){45}}

\put (276,110){\line(1,0){45}}
\put (276,112.5){\line(1,0){45}}
\put (276,107.5){\line(1,0){45}}



\put(148,140){\makebox(0,0){$X$}}
\put(148,155){\makebox(0,0){$i\gamma_\mu \gamma_5$}}




\end{picture}
\vspace{0.5in}
\caption{Schematic baryon spin diagram, with internal quark momenta $p_1, p_2, p_3$ ;  basic spin operator $i\gamma_\mu \gamma_5$ 
is inserted in line $p_1$.}

\end{figure}

\subsection{ BS Normalization of $qqq$ Wave Function}

Consider Fig 1 where the 3 internal quark lines ($1, 2, 3$) are labelled by momenta $p_1$, $p_2$ and $p_3$ respectively. and the operator 
$i\gamma_\mu \gamma_5$ is temporarily replaced by $1$ to signify probability conservation for a BS normalization calculation. This exercise is patterned closely on the lines of \cite{MiMi84}, albeit in a suitably corrected  form in which the matrix elements are $not$ factored into two parts ( as done erroneously in \cite{MiMi84}), but otherwise  maintaining its mixed symmetric [ m' m'') notation for the matrix elements for each separate d.o.f.  ( spin and flavour). Keeping track of the indices ($1, 2, 3$), the two spin matrix elements  $N', N''$   for BS normalization 
(taken between the functions (2.7))may be written in an obvious notation as, 
\begin{eqnarray}\label{3.1}
N'   &=&  N'_{1; 23} + N'_{1; 32} + N'_{2; 31} + N'_{2; 13} + N'_{3; 12} + N'_{3; 21}  \nonumber  \\
N''  &=&  N''_{1; 23} + N''_{1; 32} + N''_{2; 31} + N''_{2; 13} + N''_{3; 12} + N''_{3; 21}
\end{eqnarray}
 and should be multiplied by the corresponding flavour matrix elements (2.8) in accordance with the structure of the function $W(P)$ of (2.6) . The  individual terms in Eq (3.1) are related by permutation symmetry, and  two typical elements are given by 
\begin{equation}\label{3.2}
N'_{1; 23}  =  {\bar u}(P) P_s S_F(p_1) \{1\} S_F(p_1) P_E \frac{\gamma_5 C}{\sqrt{2}} S_F(p_2) \frac{C^{-1}\gamma_5}{\sqrt{2}} P_E S_F(p_3) P_s u(P) 
\end{equation}
\begin{equation}\label{3.3}
N''_{1; 23} = {\bar u}(P) P_s{\hat \gamma}_\rho \gamma_5  S_F(p_1) \{1\} S_F(p_1) P_E  \frac{\gamma_{\rho'} C}{\sqrt{6}} S_F(p_2) 
 \frac{C^{-1}\gamma_\rho}{\sqrt{6}} P_E S_F(p_3) \gamma_5 \gamma_{\rho'} P_s u(P) 
\end{equation}
where $P$ is the baryon 4-momentum with mass $M$ ( $P^2 = - M^2$),   and 
\begin{equation}\label{3.4}
P_s = (1 + i\gamma. s \gamma_5) / 2 ;  \quad  P_E = ( M - i\gamma. P) / 2
\end{equation}
and the normalization condition is ( c.f., \cite{MiMi84})  
\begin{equation}\label{3.5}
2 = \int d \tau [N' < \phi' \mid 1\mid \phi'> + N'' < \phi'' \mid 1\mid \phi'' >]  = \int d \tau [N' + N''] \equiv 2N 
\end{equation}
where the flavour functions $\phi', \phi''$ are defined in (2.8) and $ d \tau$ is the full measure of the internal integration variables defined by (2.4)
\begin{equation}\label{3.6}
d \tau \equiv  d^4 \xi d^4 \eta [D_{123} \phi ({\hat \xi}, {\hat \eta})]^2 
\end{equation}  
and the 3D wave function  $\phi$ and the associated denominator function $D_{123}$ are as defined in Eq (2.10). Note that the time-like variables 
$\xi_0$ and $\eta_0$ of Eq.(2.2) do not appear in the factors $D_{123} \phi$ on the rhs of (3.6).  
We may now use the same pattern for the evaluation of some standard physical quantities which may serve as checks on the self-consistency of 
this formalism. Thus for the nucleon charge, the probability operator $1$ employed for BS normalization above should be replaced by 
\begin{equation}\label{3.7}
1 \Rightarrow  e i\gamma_\mu [ 1/6 + \tau_3 / 2 ] 
\end{equation}
and the corresponding matrix elements $Q'$ and $Q''$ may be written down in the same notation and phase convention as for $N'$ and $N''$ 
above, and then divided by the total BS normalizer $N$ for correct overall normalization. The final result for the nucleon  charge, after evaluating  
the flavour matrix elements a la Eq.(2.8) is 
\begin{equation}\label{3.8}
2Q N = \int d \tau [ Q' ( 1/6 + \tau_3 /2) + Q'' ( 1/6 - \tau_3 / 6 ) ] 
\end{equation}
where $Q'$ and $Q''$ are given by Eqs (3.2) and (3.3) respectively, except for the replacement of $\{1\}$ by $ i\gamma_\mu$, and $\tau_3$ has 
the values $ \pm 1$ for  proton / neutron. The momentum integrals are involved, but if terms of order $(\xi^2, \eta^2) / M^2$ are ignored compared 
to unity in the integrands concerned, some remarkable simplifications bring out the full flavour of $SU(6)$ symmetry, albeit in a {\it relativistic} 
manner. Thus as a first check on the self-consistency of the formalism, the proton / neutron charges work out as $e$ and $0$ respectively. 

\subsection{Spin Matrix Elements in Lowest Order}

We now employ this formalism for the determination of nucleon spin in lowest order, for which the basic spin operator is 
$i \gamma_mu \gamma_5$, (as in Fig 1),  multiplied by appropriate flavour matrices. It is simplest to speak of the `axial charges' 
whose proportionality to the spin vector $s_\mu$ comes out from analogous equations to (2) and (3) of ref. \cite{Bass05},  
 with the substitution of $\{1\}$ by $i\gamma_\mu \gamma_5$ in Eqs (3.2-3) above. The flavour dependent axial charges $g_A^{(3)}$ ,    $g_A^{(8)}$
and  $g_A^{(0)}$ of ref.\cite{Bass05} are then reproduced by the multiplication of this spin operator with  the successive Gell-Mann  matrices 
$\lambda_{3, 8, 0}$ respectively, and taking their  matrix elements between  the states defined by (2.8).  Now the spin anomaly occurs mainly 
with respect to $g_A^{(0)}$, while the other two parameters remain almost unaffected. In the lowest order, i.e., neglecting terms of order 
$(\xi^2, \eta^2) / M^2$,  these quantities may be worked out in the same normalization as defined in Section 3.1 above, to yield the values 
\begin{equation}\label{3.9}
g_A^{(3)} = 10 / 9 ; \quad   g_A^{(8)} = 2 / 3;  \quad g_A^{(0)} = 2 /3 
\end{equation} 
Comparison with  Eq.(9) of ref.\cite{Bass05} reveals a difference of a factor of $2/3$ between the two results. This is due to the BS 
normalization employed here, viz., a relativistic one normalizing direct to  unit probability which does $not$ distinguish between the proton and the neutron ),  instead of to the charge  which does,  as in ref \cite{MiMi84}. The latter agrees with the standard non-relativistic value cited in  ref.\cite{Bass05}, but the former indicates a welcome alternative possibility to ensure  better with experiment without relativistic corrections. Further, it         
 is only the last one, $g_A^{(0)}$, that is subject to anomaly corrections arising mainly from two-gluon effects that we consider next.

\section{Spin correction from Two-gluon Anomaly }

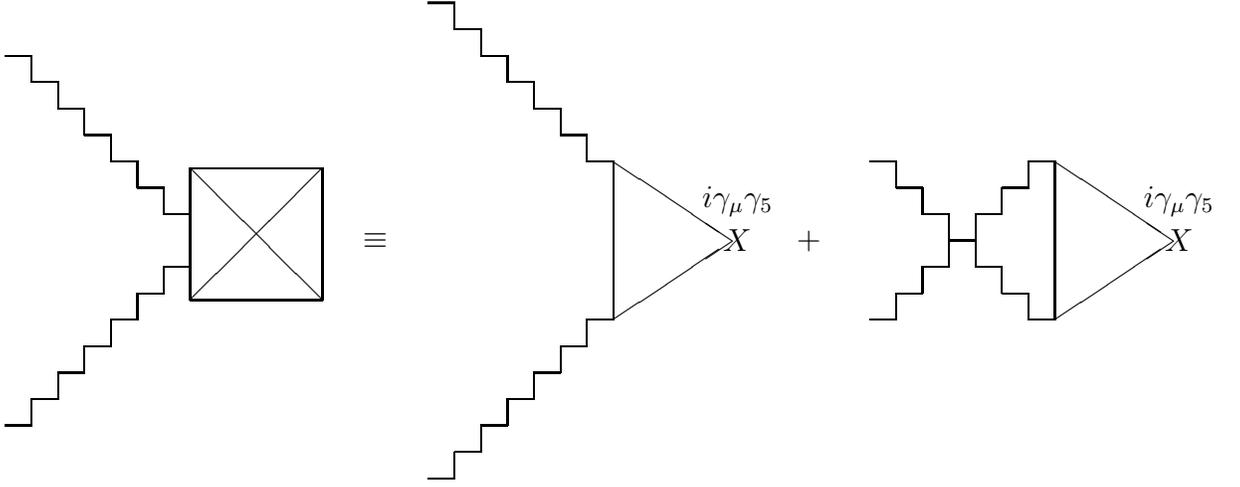
\begin{figure}[h]

\vspace{0.5in}

\begin{picture}(450,50)(-60,100)

\multiput(267,80)(10,10){7}{\line(1,0){10}}
\multiput(277,80)(10,10){6}{\line(0,1){10}}
\multiput(267,140)(10,-10){7}{\line(1,0){10}}
\multiput(277,130)(10,-10){6}{\line(0,1){10}}

\put (337,80){\line(0,1){60}}
\put (337,80){\line(3,2){45}}
\put (337,140){\line(3,-2){45}}

\put (170,80){\line(0,1){60}}
\put (170,80){\line(3,2){45}}
\put (170,140){\line(3,-2){45}}

\multiput(100,20)(10,10){7}{\line(1,0){10}}
\multiput(110,20)(10,10){6}{\line(0,1){10}}
\multiput(100,200)(10,-10){7}{\line(1,0){10}}
\multiput(110,190)(10,-10){6}{\line(0,1){10}}

\put (10,87.5){\line(0,1){50}}
\put (60,87.5){\line(0,1){50}}
\put (10,87.5){\line(1,0){50}}
\put (10,137.5){\line(1,0){50}}
\put (10,87.5){\line(1,1){50}}
\put (10,137.5){\line(1,-1){50}}

\put(80,110){\makebox(0,0){$\equiv$}}

\multiput(-60,40)(10,10){7}{\line(1,0){10}}
\multiput(-50,40)(10,10){6}{\line(0,1){10}}
\multiput(-60,180)(10,-10){7}{\line(1,0){10}}
\multiput(-50,170)(10,-10){6}{\line(0,1){10}}

\put(217,110){\makebox(0,0){$X$}}
\put(384,110){\makebox(0,0){$X$}}
\put(217,125){\makebox(0,0){$i\gamma_\mu \gamma_5$}}
\put(384,125){\makebox(0,0){$i\gamma_\mu \gamma_5$}}




\put(244,110){\makebox(0,0){$+$}}

\end{picture}
\vspace{1.25in}
\caption{Two gluon operator (crossed box) representing a sum of two distinct diagrams for axial vector coupling }

\end{figure}

\subsection{Two-gluon Anomaly Operator}

\setcounter{equation}{0}
\renewcommand{\theequation}{4.\arabic{equation}}

The 2-gluon anomaly operator $ \Delta_{\mu\nu\lambda}$ appears in Fig 2 as a `crossed box'  represented by a sum of two triangle diagrams, the second one being merely the effect of  exchanging the two gluon lines connected to the triangle loop.  In this Section we indicate its evaluation in a general manner in preparation  for its  insertion  in the internal quark lines (Fig 3) for obtaining the gluon anomaly corrections to $g_A^{(0)}$. The 2-gluon 
anomaly  operator, with gluon momenta $k1 =k$ (entering)  and $k_2 = k$ (leaving ) may be expressed in the  form 
\begin{equation}\label{4.1}
\Delta_{\mu\nu\lambda} (k) = \frac{i g_s^2}{(2\pi)^4} Tr[ \int d^4 q i\gamma_\nu S_F(q+k_1) i\gamma_\mu \gamma_5 S_F(q+k_2) i\gamma_\lambda S_F(q)] 
\end{equation}
A second one is obtained by the simultaneous interchanges $ k \rightarrow -k $ and  $\nu \rightarrow \ \lambda$. The calculation is straightforward 
and will be mostly skipped except for a quick indication of how to incorporate gauge invariance. While the modern method is that of dimensional 
regularization, it should be adequate to follow an old-fashioned (simpler) method due to Rosenberg \cite{Rose63}, which effectively amounts to subtracting out the non-gauge-invariant terms at the integrand itself,  so as to ensure separate conservation of currents at the two vertices $\nu$ and $\lambda$.   
After the trace evaluation in (3.9), this procedure leaves a numerator proportional to $q$ in the integrand. This needs at least an extra power of $q$ arising 
from an expansion of the propagator denominators in powers of $q.k /(q^2 +k^2)$.  In the lowest order in $k$,  the integral over $q^2$ becomes convergent, and after standard $q$ integration via the Feynman auxiliary variable $u$, reduces to an integral over $u$ 
$$ \Delta_{\mu\nu\lambda} \approx \frac{2 \alpha_s}{\pi} \int_0^1 du u^2 / [m_q^2 + k^2 u] $$ 
which for small  $m_q^2 $  further reduces to a very simple form :
\begin{equation}\label{4.2}
\Delta_{\mu\nu\lambda} \approx \alpha_s \epsilon_{\mu\nu\lambda\sigma} k_\sigma; \quad m_q^2 << k^2 
\end{equation}    

\subsection{  2-gluon anomaly correction to spin amplitude}

\begin{figure}[h]

\vspace{0.5in}

\begin{picture}(450,50)(-80,100)

\put (276,80){\line(0,1){60}}
\put (276,80){\line(-3,2){45}}
\put (231,110){\line(3,2){45}}

\put (20,80){\line(0,1){60}}
\put (65,110){\line(-3,2){45}}
\put (65,110){\line(-3,-2){45}}

\put (20,80){\line(1,0){256}}
\put (20,140){\line(1,0){256}}
\put (65,110){\line(1,0){166}}

\put(75,85){\makebox(0,0){$p_3$}}
\put(75,145){\makebox(0,0){$p_1$}}
\put(75,115){\makebox(0,0){$p_2$}}

\put (-25,110){\line(1,0){45}}
\put (-25,112.5){\line(1,0){45}}
\put (-25,107.5){\line(1,0){45}}

\put (276,110){\line(1,0){45}}
\put (276,112.5){\line(1,0){45}}
\put (276,107.5){\line(1,0){45}}







\put (276,-40){\line(0,1){60}}
\put (276,-40){\line(-3,2){45}}
\put (231,-10){\line(3,2){45}}

\put (20,-40){\line(0,1){60}}
\put (65,-10){\line(-3,2){45}}
\put (65,-10){\line(-3,-2){45}}

\put (20,-40){\line(1,0){256}}
\put (20,20){\line(1,0){256}}
\put (65,-10){\line(1,0){166}}

\put(75,-35){\makebox(0,0){$p_3$}}
\put(75,25){\makebox(0,0){$p_1$}}
\put(75,-5){\makebox(0,0){$p_2$}}

\put (-25,-10){\line(1,0){45}}
\put (-25,-12.5){\line(1,0){45}}
\put (-25,-7.5){\line(1,0){45}}

\put (276,-10){\line(1,0){45}}
\put (276,-12.5){\line(1,0){45}}
\put (276,-7.5){\line(1,0){45}}

\put (154,-30){\line(0,1){10}}
\put (154,-20){\line(1,0){10}}
\put (154,-30){\line(1,0){10}}
\put (164,-30){\line(0,1){10}}
\put (154,-30){\line(1,1){10}}
\put (154,-20){\line(1,-1){10}}


\put (137,154){\line(0,1){10}}
\put (137,164){\line(1,0){10}}
\put (137,154){\line(1,0){10}}
\put (147,154){\line(0,1){10}}
\put (137,154){\line(1,1){10}}
\put (137,164){\line(1,-1){10}}

\multiput(128,142)(2,2){7}{\line(1,0){2}}
\multiput(128,140)(2,2){7}{\line(0,1){2}}
\multiput(142,154)(2,-2){7}{\line(1,0){2}}
\multiput(144,152)(2,-2){7}{\line(0,1){2}}

\multiput(140,-40)(2,2){7}{\line(1,0){2}}
\multiput(142,-40)(2,2){6}{\line(0,1){2}}
\multiput(140,-10)(2,-2){7}{\line(1,0){2}}
\multiput(142,-12)(2,-2){6}{\line(0,1){2}}

\put(142,60){\makebox(0,0){$(a)$}}
\put(142,-60){\makebox(0,0){$(b)
$}}

\end{picture}
\vspace{2.5in}
\caption{Two-gluon operator, fig (2), inserted in the internal quark lines of the baryon: (a) `self-energy' like insertion in line $p_1$; 
(b) `exchange-like' insertion connecting  lines $p_2$ and $p_3$  }

\end{figure}
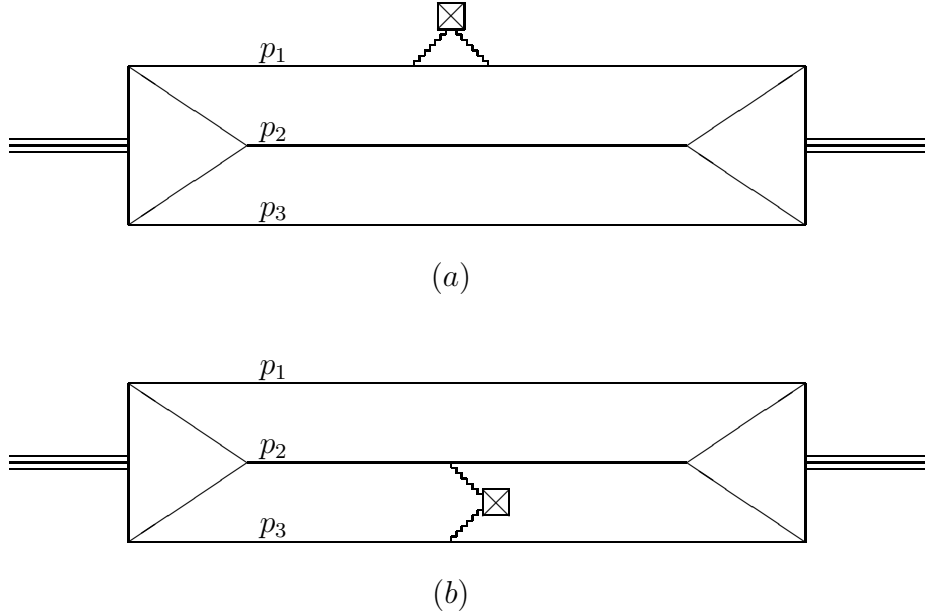

The operator $ \Delta_{\mu\nu\lambda}$  is now ready for insertion in the internal quark lines of Fig. 3 signifying the forward scattering 
amplitude of the baryon. The insertion can be done in two different ways :  self-energy like insertion in line $p_1$ a la Fig.3 (a); and 
exchange like insertion connecting two quark lines $p_1$ and $p_2$, as in Fig 3(b). We designate these contributions by $\Sigma'$ ,
$\Sigma''$ ; and $V'$, $V''$  respectively, in accordance with the two types of spin matrix elements a la Eq.(2.7). These contributions are further 
indexed by the subscripts $1; 23$, etc since  three such diagrams for each type  must be added up like in Eq.(3.1). The master expressions 
for these matrix elements are as follows. 
\begin{eqnarray}\label{4.3}
\Sigma'_{1;23}   &=& \frac{2 g_s^2}{3 (2\pi)^4} \int d^4 k {\bar u}(P) P_s S_F(p_1)  \Delta_{\mu\nu\lambda}i\gamma_\nu S_F(p_1- k)i\gamma_\lambda D^2 (k)
\nonumber \\
                           & &   S_F(p_1) P_E \frac{ \gamma_5 C}{\sqrt{2}}S_F(-p_2) \frac{C^{-1}  \gamma_5}{\sqrt{2}} P_E S_F(p_3) P_s u(P) + conj 
\end{eqnarray}
\begin{eqnarray}\label{4.4}
\Sigma''_{1;23}  &=& \frac{2 g_s^2}{3 (2\pi)^4} \int d^4 k {\bar u}(P) P_s{\hat \gamma}_\rho \gamma_5  S_F(p_1)  \Delta_{\mu\nu\lambda}i\gamma_\nu S_F(p_1- k) i\gamma_\lambda D^2 (k)
\nonumber \\
                           & &  S_F(p_1) P_E \frac{i{\hat \gamma}_{\rho'}C}{\sqrt{6}} S_F(-p_2) \frac{C^{-1} i{\hat  \gamma}_\rho}{\sqrt{6}} P_E S_F(p_3) P_s u(P) + conj 
\end{eqnarray}
The symbols $conj$ in these equations represent the effects of the crossed diagrams for the 2-gluon anomaly (Fig 2). 
For the exchange type insertions, the corresponding expressions $V', V''$ may be written in a similar but slightly simplified form as   
\begin{eqnarray}\label{4.5}
V'_{1;23}    &=&   \frac{2 g_s^2}{3 (2\pi)^4} \int d^4 k {\bar u}(P) P_s S_F(p_1)\frac{P_E \gamma_5}{2} \Delta_{\mu\nu\lambda}S_F(-p_2 + k) \nonumber \\
                   & & \gamma_\nu D(k) S_F(-p_2) \gamma_5 P_E S_F(p_3) \gamma_\lambda D(k) \nonumber \\
                   & &  \times S_F (p_3 + k) P_s u(P) + conj 
\end{eqnarray}
 \begin{eqnarray}\label{4.6}
V''_{1;23}   &=&   \frac{2 g_s^2}{3 (2\pi)^4} \int d^4 k {\bar u}(P) P_s {\hat \gamma}_\rho \gamma_5 P_E S_F(p_1)\frac{P_E{\hat \gamma}_{\rho'}}{6} \Delta_{\mu\nu\lambda}S_F(-p_2 + k)\nonumber \\ 
                   & &\gamma_\nu D(k) S_F(-p_2){\hat  \gamma}_\rho  P_E S_F(p_3) \gamma_\lambda D(k)  \nonumber  \\
                   & & \times S_F (p_3 + k) \gamma_5 {\hat \gamma}_{\rho'}  u(P) + conj 
 \end{eqnarray}

These quantities, when integrated over $ \int d\tau$, Eq. (3.6),  and divided by the normalizer $N$, Eq.(3.5),  qualify directly as 2-gluon anomaly corrections (in the same relative normalization) 
to the spin matrix element $g_A^{(0)} $ listed in   (3.9). The result for the fractional correction to $g_A^{(0)}$ \\ 
may be expressed  in the form 
\begin{equation}\label{4.7}
\delta g_A = \theta [\frac{\alpha_s}{\pi}]^2 g_A^{(0)}.
\end{equation}
where the dimensionless quantity $\theta$ may be termed  the `reduced fractional   2-gluon anomaly  correction' . 
\par
 The calculation of $\theta$ -  a long and elaborate proces -   involves two distinct steps : 
(a) integration over $d^4 k$   (b) integration  over $d \tau$. While step (a) is necessarily a dynamic correction,  step (b)  may be further divided into 
two parts, i) `kinematic'  and ii) `dynamic',  according as the effects of the internal momenta ($\xi, \eta$) are neglected or included respectively.  The 
reason for this break -up is that only the latter  involves an interplay of the the 3D wave function $[\mid \phi \mid]^2$, appearing via the integration measure $ d\tau$, with the internal momenta ($\xi, \eta$) which are copiously present in  the large number of propagators which make up the integrands 
of the types (4.3 - 4.6) , while the `kinematical' part almost entirely suppresses this contribution by dropping the effects of these internal momenta from 
the said propagators. [Note that the hypergeometric form (2.15) of $\phi$ which appears through the integral measure $d \tau$,  carries  the $dynamical$ 
signature of the `spin-part' of the 3-body force !].  In this paper we are able to  give only the results of the `kinematical' part,  while   the  calculation of the more difficult `dynamical' part is  in progress.  To that end, the `kinematical'  part is  calculable on closely analogous lines to the  spin matrix elements in lowest  order (see Sect. 3.2), using the normalization of Sect (3.1). The essential steps are very briefly indicated below.  
 
\subsection{ `Kinematical' Part of the Spin Correction} 

First, to incorporate the operator $\Delta_{\mu\nu\lambda}$ of Eq. (3.10), the following results are useful:
\begin{equation}\label{4.8}
\gamma_\nu \gamma_\lambda \gamma_\sigma \epsilon_{\mu\nu\lambda\sigma} = 6 \gamma_\mu \gamma_5 ; \quad 
 \gamma_\lambda \gamma_\sigma \epsilon_{\mu\nu\lambda\sigma} = 2 \gamma_\nu \gamma_\mu \gamma_5 
\end{equation}
Next, the (logarithmic) divergence of the $k$- integration  requires the standard process of dimensional regularization
\cite{tHVe72}, with a typical result of the form \cite{MiHw05} 
\begin{equation}\label{4.9}
\int \frac{d^4k}{i} \int_0^1 du 2(1-u) \frac{ k^2}{(k^2 + \Lambda_u)^3} = - \pi^2 [ \gamma - 1 + \ln{\pi \Delta_1}]
\end{equation}
where  
$$ \Lambda_u = u \Delta_1 + m_g^2 (1-u); \quad \Delta_1 = m_q^2 + p_1^2 $$ 
After the $k$-integration (step (a)), the $d\tau$ integration (step (b)) involves some drastic approximations effectively  involving the replacement of the 
4-momenta $p_i$ of the various propagators by their `central' values. At the end of this exercise, the effect of the factor $\phi^2$ in $d\tau$ almost `decouples' from that of the various propagators involved in step (b), and the integrations can be performed without much further ado. Omitting these
steps, the  two contributions $\theta_1$ and $\theta_2$ from the `self-energy' and `exchange' effects respectively become the following :
\begin{equation}\label{4.10}
\theta_1 \approx - 0.5 ;  \quad  \theta_2 \approx - 1.5   
\end{equation} 
resulting in a total effect `kinematical' contribution 
\begin{equation}\label{4.11} 
 \theta \approx  - 2.0 
\end{equation}  
which with $ \alpha \approx 0.39$ in (4.7),  amounts to a tiny correction to the spin anomaly, albeit of the $right$  sign.   

\section{Summary and Conclusion}

To summarise, we have presented a first application of a new form of dynamics 
within the framework of QCD in the high momentum limit, viz., the role of a direct 
$qqq$ force which has been shown \cite{Mitr07} to produce an additional singularity in the 
structure $\phi$  of the 3D  $qqq$ wave function. The application is intended to address the  
issue of the proton spin anomaly in terms of a two-gluon anomaly effect. 
To that end, a good part of the paper has been devoted to a fairly general formulation of baryonic transition amplitudes, 
looked upon as $qqq$ systems in terms of Feynman amplitudes involving appropriate quark loops.  The Bethe-Salpeter 
normalization has been attuned to the total probability which maintains a symmetry between the proton and the neutron, 
instead of to the  total charge which does not. This relativistic formulation  has the advantage that the axial charges $g_A^{(i)}$, ($ i = 0,8,3$), are 
already $2/3$ times the corresponding non-relativistic quantities \cite{Bass05},  thus obviating  major `relativistic corrections' \cite{Bass05} for them. 
Thus calibrated, the formalism is applied to the evaluation of two-gluon anomaly corrections [self-energy and exchange]  to  $g_A^{(0)}$, by inserting the anomaly operator $\Delta_{\mu\nu\lambda}$ into the internal quark lines, so as to produce a fractional correction of the general form (4.7), in which 
the dimensionless quantity $\theta$ is a measure of the correction.  Unfortunately  we have so far been able to calculate only the `kinematical' correction which corresponds to the neglect of the internal momenta ($\xi, \eta$)  in the integrands of the amplitudes involved. The resulting value of 
$\theta$ is $- 2.0$ which  has the right sign, but  a rather small magnitude. This still leaves open the possibilities of  `dynamical' corrections which involve an interplay of the internal momenta, mostly arising from the various propagators, with the 3D wave function $\phi$ whose hypergeometric 
form (2.14)  reflects the dynamics of the 3-body force, namely the   $negative$  eigenvalue of the associated spin operator.   The `correct' 
(negative) sign of $\theta$ is an encouraging sign for the  vast scope for  the role of this crucial dynamics  yet to be included in its derivation. 
This calculation is currently in progress. 
\par
The author is grateful to the organizers of THEOPHYS-07 for an opportunity to present these preliminary results at this Conference.

\end{document}